\documentclass[article,12pt]{elsarticle}
\usepackage{a4wide}
\usepackage{natbib} 			
\usepackage{graphicx}
\usepackage{amsfonts}
\usepackage{amsmath}
\usepackage{booktabs}
\usepackage{epsfig}
\usepackage[colorlinks]{hyperref}
\hypersetup{citecolor=black,linkcolor= black}

\begin{document}
\title{Detrending moving-average cross-correlation coefficient: Measuring cross-correlations between non-stationary series}
\author{Ladislav Kristoufek}
\ead{kristouf@utia.cas.cz}
\address{Institute of Information Theory and Automation, Academy of Sciences of the Czech Republic, Pod Vodarenskou Vezi 4, 182 08, Prague 8, Czech Republic\\
Institute of Economic Studies, Faculty of Social Sciences, Charles University, Opletalova 26, 110 00, Prague 1, Czech Republic
}

\begin{abstract}
In the paper, we introduce a new measure of correlation between possibly non-stationary series. As the measure is based on the detrending moving-average cross-correlation analysis (DMCA), we label it as the DMCA coefficient $\rho_{DMCA}(\lambda)$ with a moving average window length $\lambda$. We analytically show that the coefficient ranges between -1 and 1 as a standard correlation does. In the simulation study, we show that the values of $\rho_{DMCA}(\lambda)$ very well correspond to the true correlation between the analyzed series regardless the (non-)stationarity level. Dependence of the newly proposed measure on other parameters -- correlation level, moving average window length and time series length -- is discussed as well.
\end{abstract}


\journal{Physica A}

\maketitle


\section{Introduction}

Inspection of statistical properties of multivariate series has become a topic of increasing importance in econophysics. For this purpose, various estimators of power-laws in cross-correlations of a pair of series have been proposed \citep{Podobnik2008,Zhou2008,Kristoufek2011,He2011a,Wang2012}. Out of these, the detrended cross-correlation analysis (DCCA) \citep{Podobnik2008,Zhou2008} has become the most popular one. Apart from the analysis of the power laws in the cross-correlation function itself, Zebende \citep{Zebende2011} proposed the DCCA cross-correlation coefficient as a combination of DCCA and the detrended fluctuation analysis (DFA) \cite{Peng1993,Peng1994,Kantelhardt2002}. Even though its ability to uncover power-law cross-correlations has been somewhat disputed \citep{Podobnik2011,Balocchi2013,Blythe2013,Zebende2013}, Kristoufek \citep{Kristoufek2013b} shows that the coefficient is able to estimate the correlation coefficient between non-stationary series precisely and that it dominates the standardly used Pearson's correlation coefficient.

In this paper, we propose an alternative but also a complementary coefficient to the DCCA cross-correlation coefficient -- detrending moving-average cross-correlation coefficient. In the following section, the coefficient is introduced. After that, results of the wide Monte Carlo study are presented showing that the newly proposed coefficient estimates the true correlation coefficient precisely even for strongly non-stationary series. 

\section{DMCA coefficient}

We start with the detrending moving average procedure (DMA) proposed by Vandewalle \& Ausloos \cite{Vandewalle1998} and further developed by Alessio \textit{et al.} \cite{Alessio2002}. For (possibly asymptotically non-stationary) series $\{x_t\}$ and $\{y_t\}$, we construct integrated series $X_t=\sum_{i=1}^{t}{x_i}$ and $Y_t=\sum_{i=1}^{t}{y_i}$ for $t=1,2,\ldots,T$ where $T$ is the time series length which is common for both series. Fluctuation functions $F_{x,DMA}$ and $F_{y,DMA}$ are then defined as
\begin{equation}
F_{x,DMA}^2(\lambda)=\frac{1}{T-\lambda+1}\sum_{i=\lfloor \lambda-\theta(\lambda-1)\rfloor}^{\lfloor T-\theta(\lambda-1)\rfloor}{\left(X_t-\widetilde{X_{t,\lambda}}\right)^2},
\end{equation}
\begin{equation}
F_{y,DMA}^2(\lambda)=\frac{1}{T-\lambda+1}\sum_{i=\lfloor \lambda-\theta(\lambda-1)\rfloor}^{\lfloor T-\theta(\lambda-1)\rfloor}{\left(Y_t-\widetilde{Y_{t,\lambda}}\right)^2}
\end{equation}
where $\lambda$ is the moving average window length and $\theta$ is a factor of moving average type (forward, centered and backward for $\theta=0$, $\theta=0.5$ and $\theta=1$, respectively). $\widetilde{X_{t,\lambda}}$ and $\widetilde{Y_{t,\lambda}}$ then represent the specific moving averages with the window size $\lambda$ at time $t$. Properties of DMA have been frequently studied and the procedure is usually compared to DFA. Both methods perform similarly but DMA is much less computationally demanding as the procedure contains no box-splitting and regression fitting \cite{Carbone2003,Xu2005,Grech2005}. In a similar manner, different types of moving averages have been studied and the centered one ($\theta=0.5$) shows the best results \cite{Carbone2003}. Therefore, we apply $\theta=0.5$ in this study as well.

For the bivariate series, He \& Chen \cite{He2011a} propose the detrending moving-average cross-correlation analysis (DMCA) which can be seen as a detrended covariance. The bivariate fluctuation $F_{DMCA}^2$ is defined as
\begin{equation}
F_{DMCA}^2(\lambda)=\frac{1}{T-\lambda+1}\sum_{i=\lfloor \lambda-\theta(\lambda-1)\rfloor}^{\lfloor T-\theta(\lambda-1)\rfloor}{\left(X_t-\widetilde{X_{t,\lambda}}\right)\left(Y_t-\widetilde{Y_{t,\lambda}}\right)}.
\end{equation}
In the steps of Zebende \cite{Zebende2011}, we propose the detrending moving-average cross-correlation coefficient, or also the DMCA-based correlation coefficient, as
\begin{equation}
\rho_{DMCA}(\lambda)=\frac{F_{DMCA}^2(\lambda)}{F_{x,DMA}(\lambda)F_{y,DMA}(\lambda)}
\end{equation}

The coefficient can be rewritten as
\begin{multline}
\rho_{DMCA}(\lambda)=\frac{F_{DMCA}^2(\lambda)}{F_{x,DMA}(\lambda)F_{y,DMA}(\lambda)}=\\
\frac{\frac{1}{T-\lambda+1}\sum_{i=\lfloor \lambda-\theta(\lambda-1)\rfloor}^{\lfloor T-\theta(\lambda-1)\rfloor}{\left(X_t-\widetilde{X_{t,\lambda}}\right)\left(Y_t-\widetilde{Y_{t,\lambda}}\right)}}{\sqrt{\frac{1}{T-\lambda+1}\sum_{i=\lfloor \lambda-\theta(\lambda-1)\rfloor}^{\lfloor T-\theta(\lambda-1)\rfloor}{\left(X_t-\widetilde{X_{t,\lambda}}\right)^2}\frac{1}{T-\lambda+1}\sum_{i=\lfloor \lambda-\theta(\lambda-1)\rfloor}^{\lfloor T-\theta(\lambda-1)\rfloor}{\left(Y_t-\widetilde{Y_{t,\lambda}}\right)^2}}}=\\
\frac{\sum_{i=\lfloor \lambda-\theta(\lambda-1)\rfloor}^{\lfloor T-\theta(\lambda-1)\rfloor}{\epsilon_{x,t}\epsilon_{y,t}}}{\sqrt{\sum_{i=\lfloor \lambda-\theta(\lambda-1)\rfloor}^{\lfloor T-\theta(\lambda-1)\rfloor}{\epsilon_{x,t}^2}\sum_{i=\lfloor \lambda-\theta(\lambda-1)\rfloor}^{\lfloor T-\theta(\lambda-1)\rfloor}{\epsilon_{y,t}^2}}}
\label{DMCA}
\end{multline}
where $\{\epsilon_{x,t}\}$ and $\{\epsilon_{y,t}\}$ are the series $\{X_t\}$ and $\{Y_t\}$, respectively, detrended by the centered moving average of length $\lambda$. From the last part of Eq. \ref{DMCA}, it is visible that
\begin{equation}
-1\le \rho_{DMCA}(\lambda) \le 1
\end{equation}
according to the Cauchy-Schwartz inequality. The DMCA-based correlation coefficient thus has the same range as the standard correlation. In the following section, we show that the values of the newly proposed coefficient very precisely describe the correlations between two series and this is true even for strongly non-stationary series.


\section{Simulations results}

In this section, we show that the DMCA coefficient is able to describe correlations between (even non-stationary) series very precisely. To do so, we present a wide Monte Carlo simulation study\footnote{The R-project codes for the DMCA coefficient are available at \url{http://staff.utia.cas.cz/kristoufek/Ladislav_Kristoufek/Codes.html} or upon request from the author.} for varying level of correlations and (non-)stationarity. For this purpose, we utilize two ARFIMA(0,$d$,0) processes with correlated error terms
\begin{gather}
\label{eq:ARFIMA1}
x_t=\sum_{n=0}^{\infty}{a_n(d_1)\varepsilon_{t-n}}\\
y_t=\sum_{n=0}^{\infty}{a_n(d_2)\nu_{t-n}}
\end{gather}
where $a_n(d_i)=\frac{\Gamma(n+d_i)}{\Gamma(n+1)\Gamma(d_i)}$, $\langle \varepsilon_t \rangle=\langle \nu_t \rangle=0$, $\langle \varepsilon^2_t \rangle=\langle \nu^2_t \rangle=1$ and $\langle \varepsilon_t\nu_t \rangle=\rho_{\varepsilon\nu}$.

Parameter $d$ is important for the stationarity discussion. For $d<0.5$, the series are stationary while for $0.5 \le d < 1$, the series are non-stationary but mean-reverting whereas for $d \ge 1$, the series are non-stationary and non-mean-reverting (explosive). To see how the DMCA coefficient is able to quantify the correlation for (non-)stationary series, we select several levels of $d$, specifically $d_1=d_2 \equiv d=0.1,0.4,0.6,0.9,1.1,1.4$. To cover a wide spectrum of possible correlation levels in a sufficient detail, we study $\rho_{\varepsilon\nu}=-0.9,-0.8,\ldots,0.8,0.9$. To see the effect of different moving average lengths, we study the cases $\lambda=5,15,31,101$. And finally, we examine two time series lengths which are representative for usually analyzed series in econophysics -- $T=1000,5000$.

\begin{figure}[!htbp]
\begin{center}
\begin{tabular}{ccc}
\includegraphics[width=45mm]{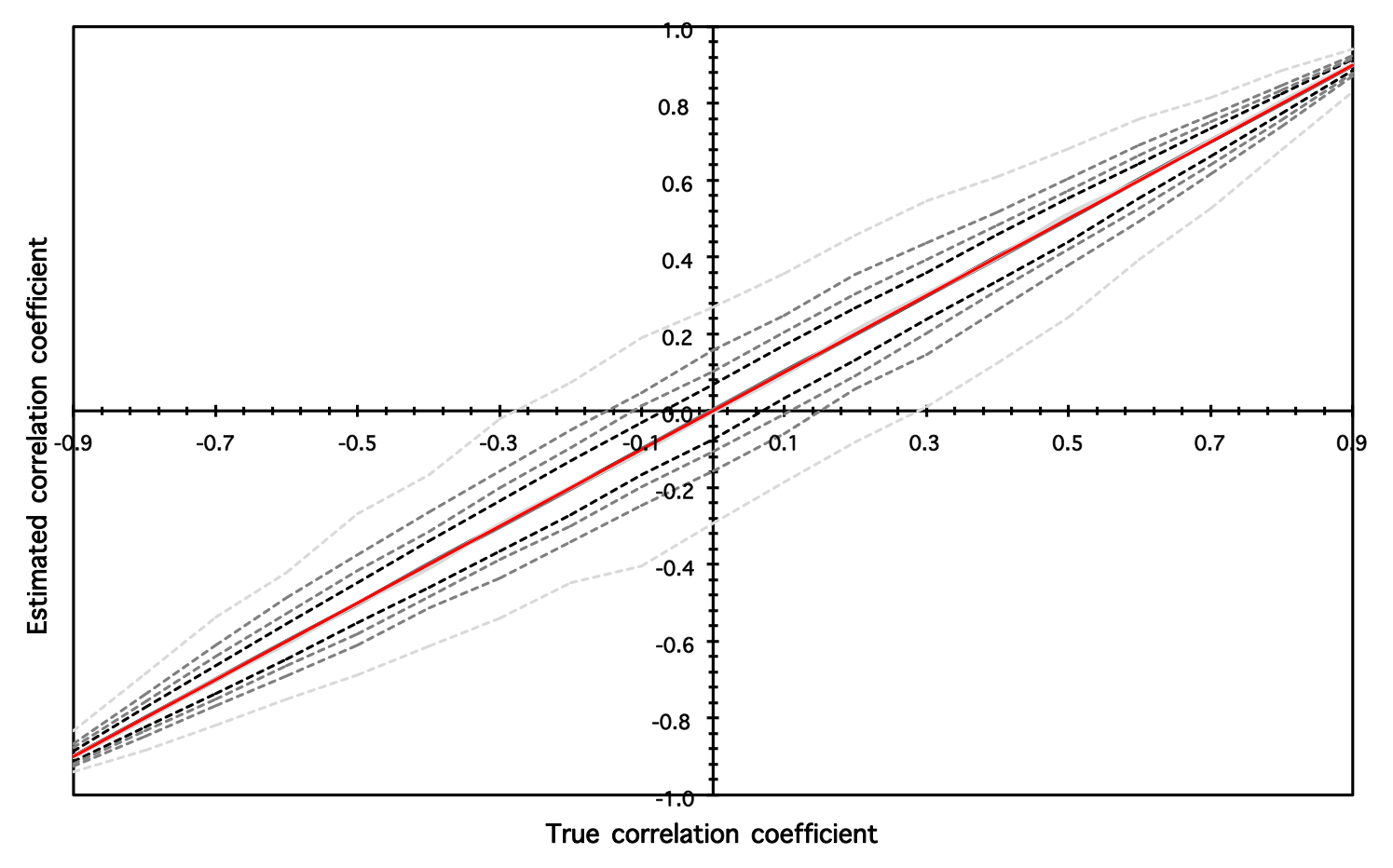}&\includegraphics[width=45mm]{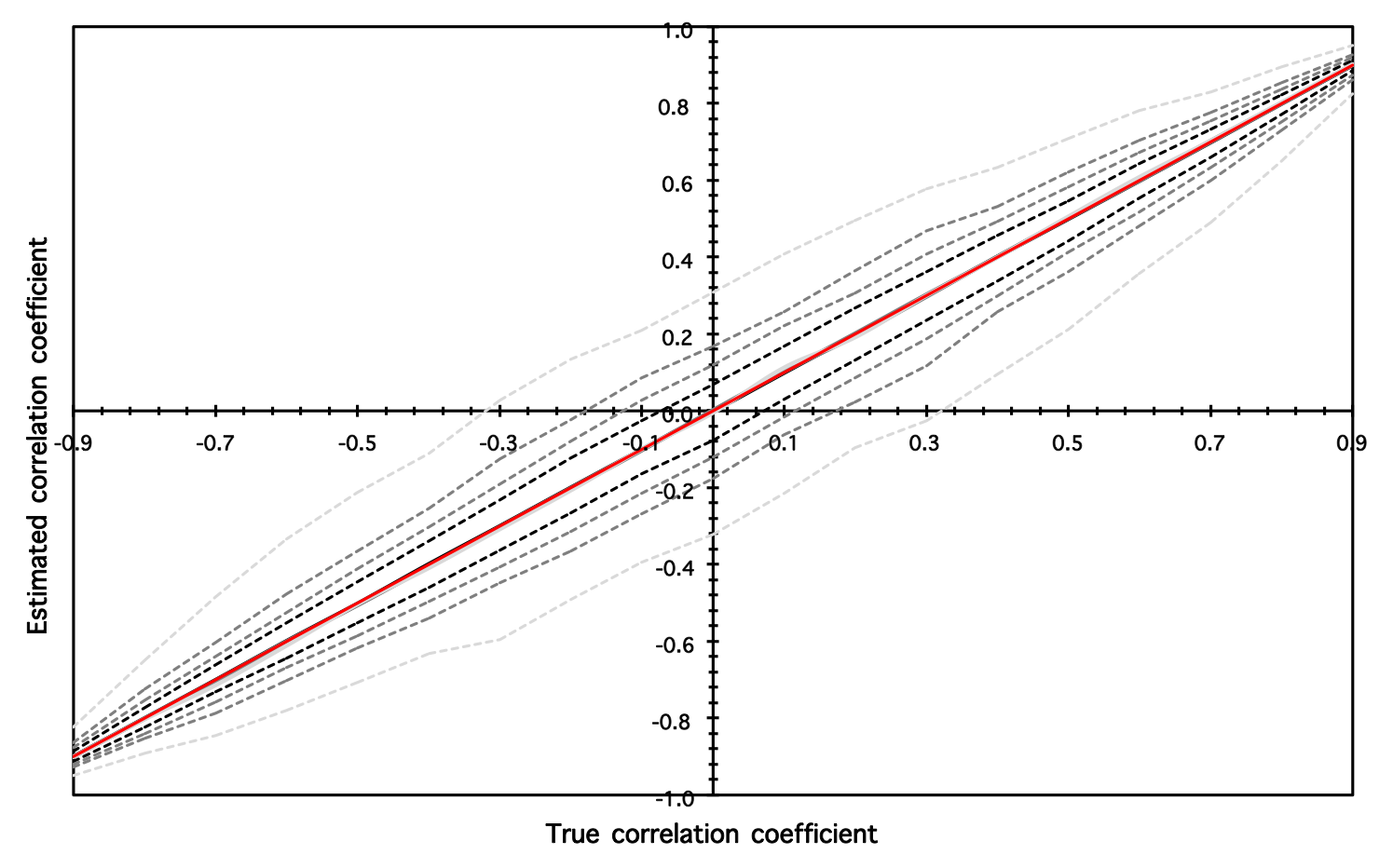}&\includegraphics[width=45mm]{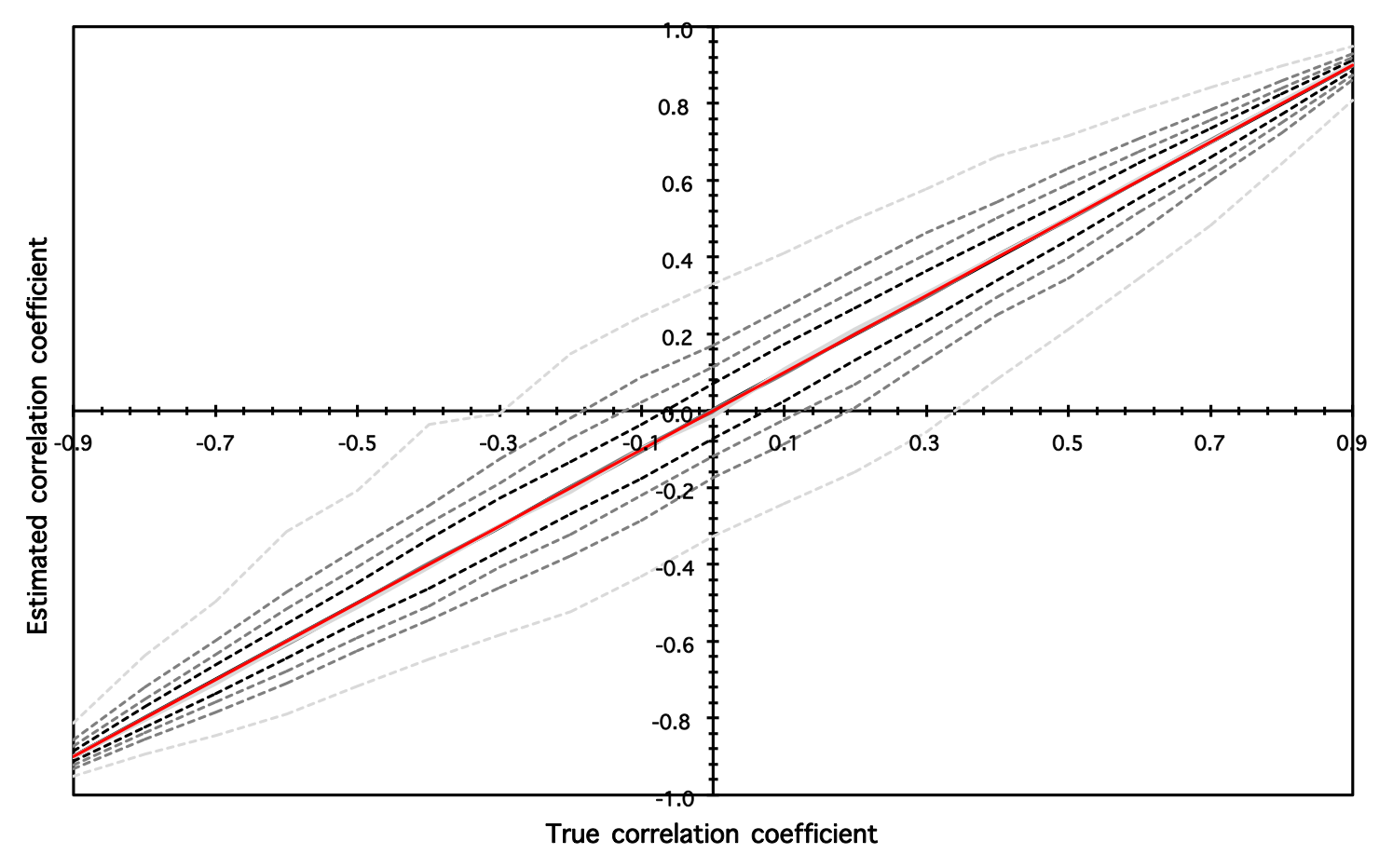}\\
\includegraphics[width=45mm]{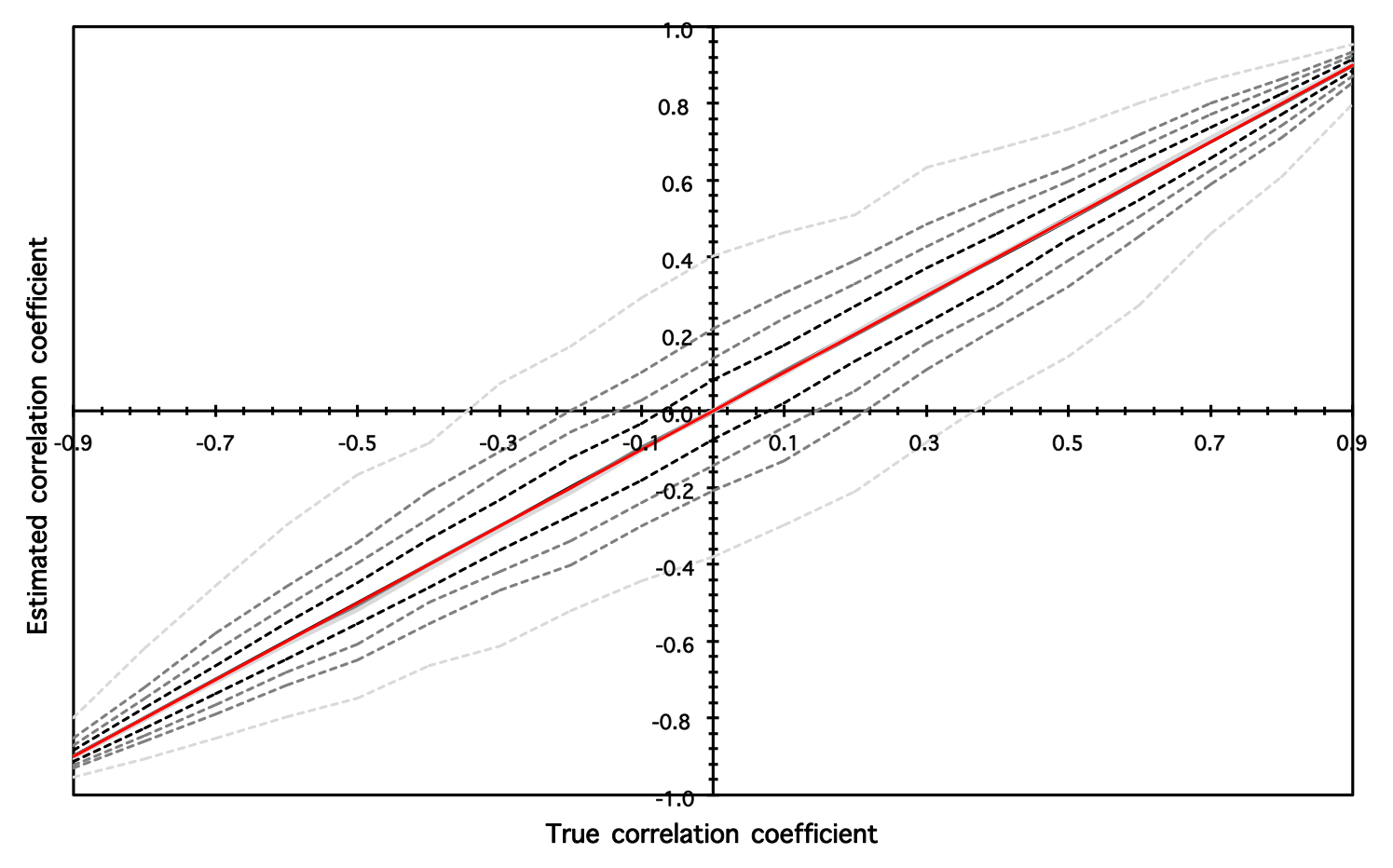}&\includegraphics[width=45mm]{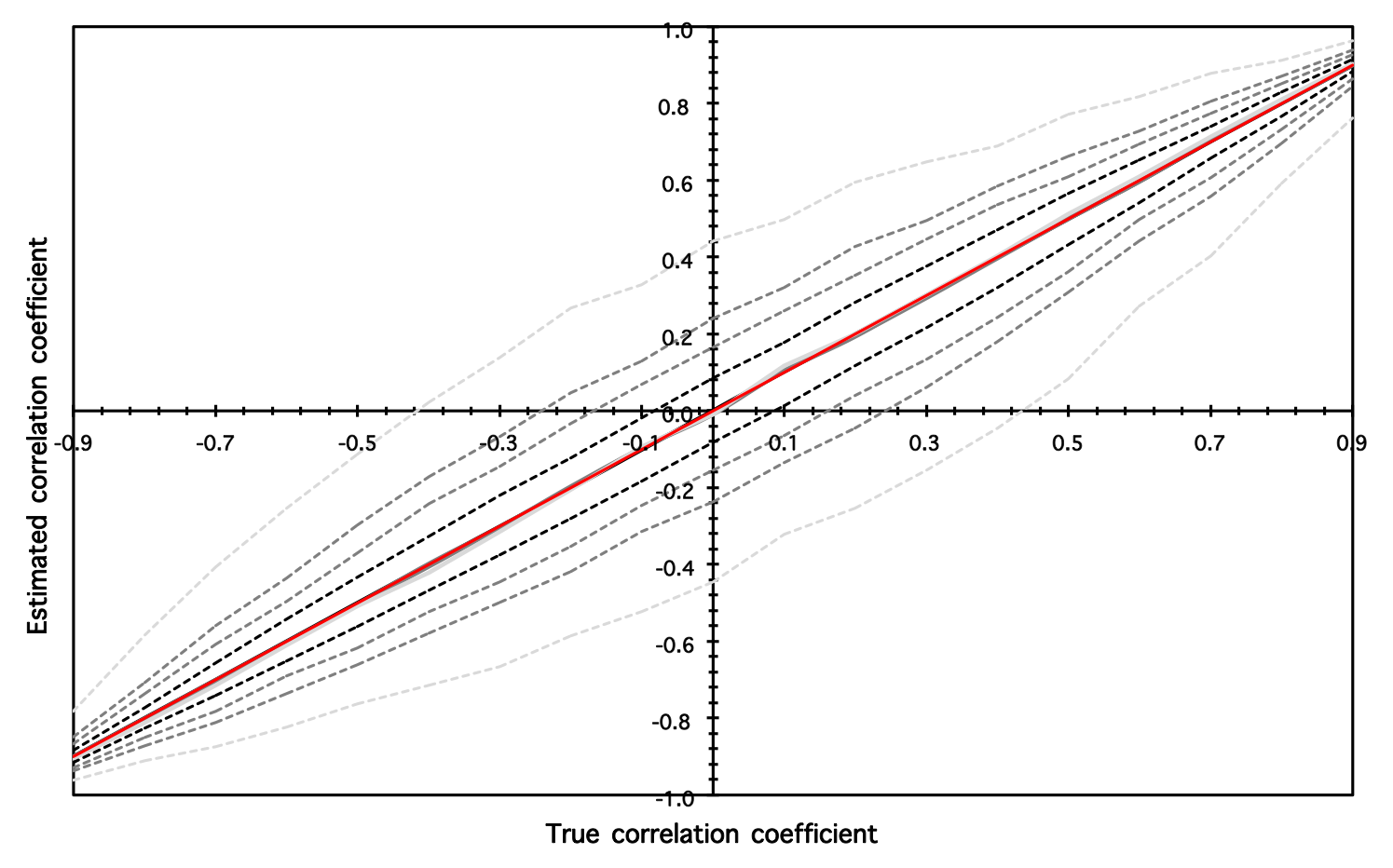}&\includegraphics[width=45mm]{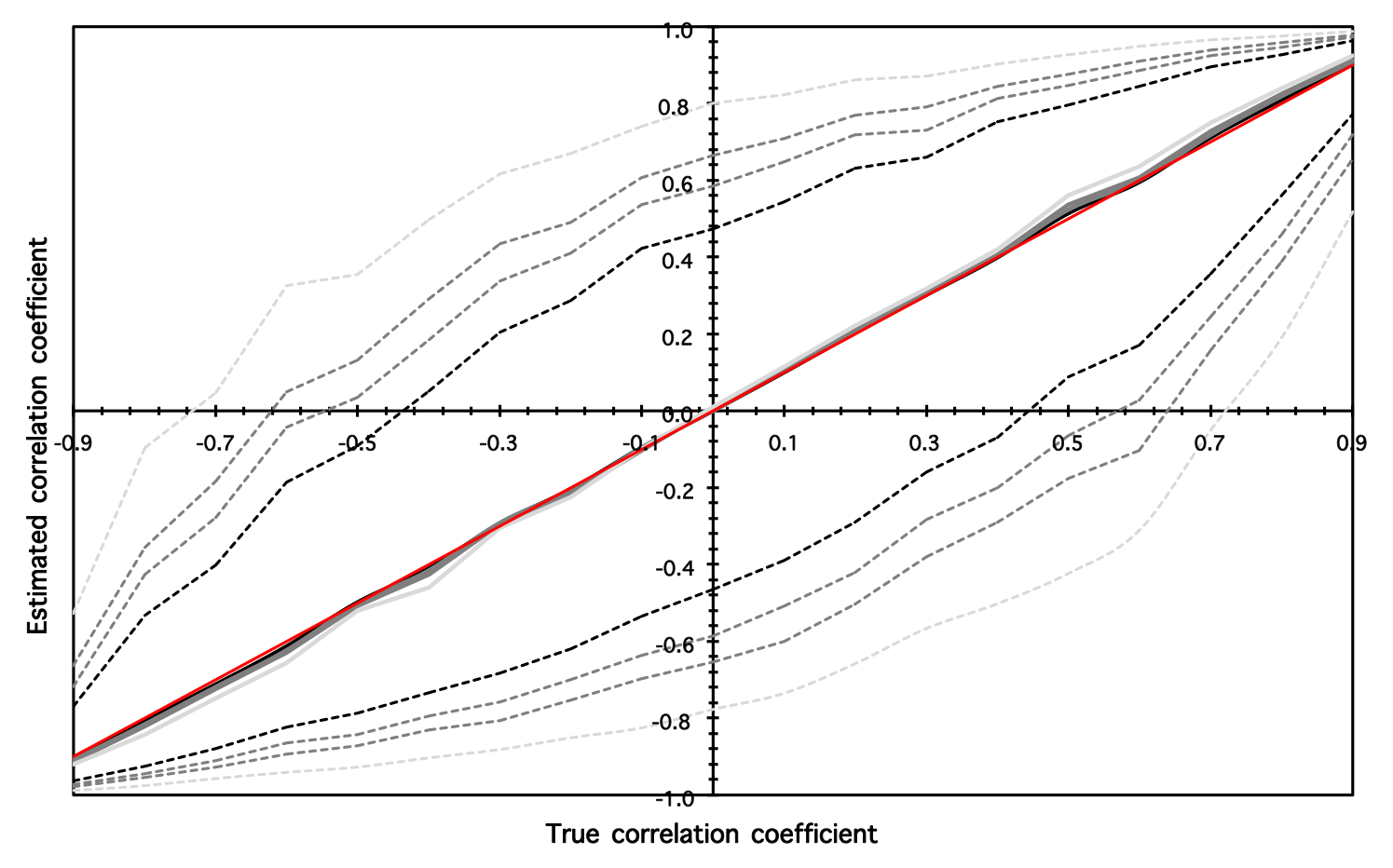}\\
\end{tabular}
\caption{\textbf{Estimated DMCA correlation coefficients for different fractional integration parameters $d$ I.} \footnotesize{Results for the time series of length $T=1000$ are shown here. Separate figures represent different parameters $d$ -- $d=0.1$ (top left), $d=0.4$ (top center), $d=0.6$ (top right), $d=0.9$ (bottom left), $d=1.1$ (bottom center), $d=1.4$ (bottom right). Red lines represent the true value of $\rho_{\varepsilon\nu}$. The solid lines of shades of grey (mostly overlapping with the red line) represent the median values of 1000 simulations for the given parameter setting. The dashed lines represent the 95\% confidence intervals (the 2.5th and the 97.5th quantiles of the simulations). Different shades of grey stand for different values of the moving average window $\lambda$ going from the lowest one ($\lambda=5$, the darkest shade) to the highest one ($\lambda=101$, the lightest shade).}\label{fig1}}
\end{center}
\end{figure}

\begin{figure}[!htbp]
\begin{center}
\begin{tabular}{ccc}
\includegraphics[width=45mm]{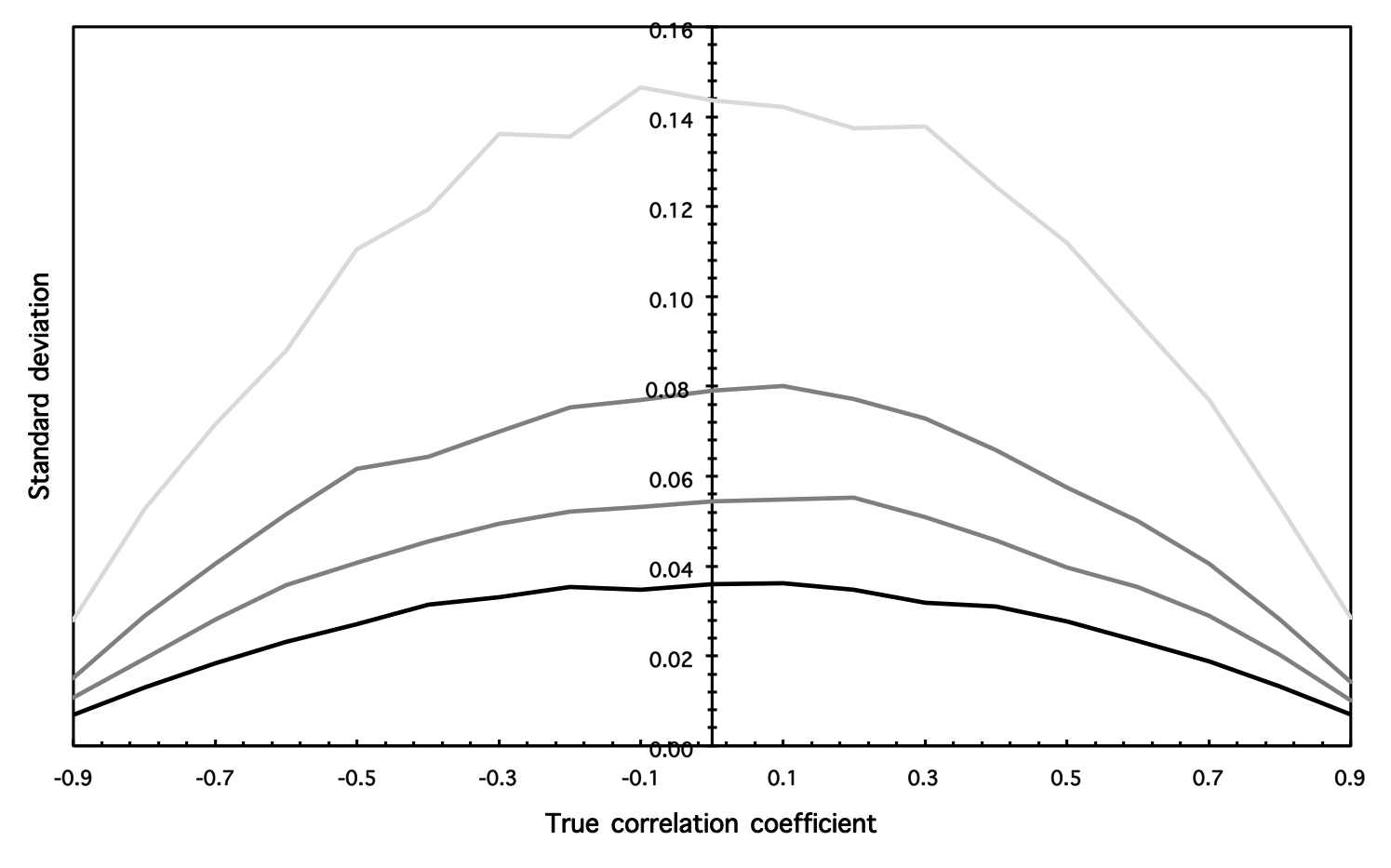}&\includegraphics[width=45mm]{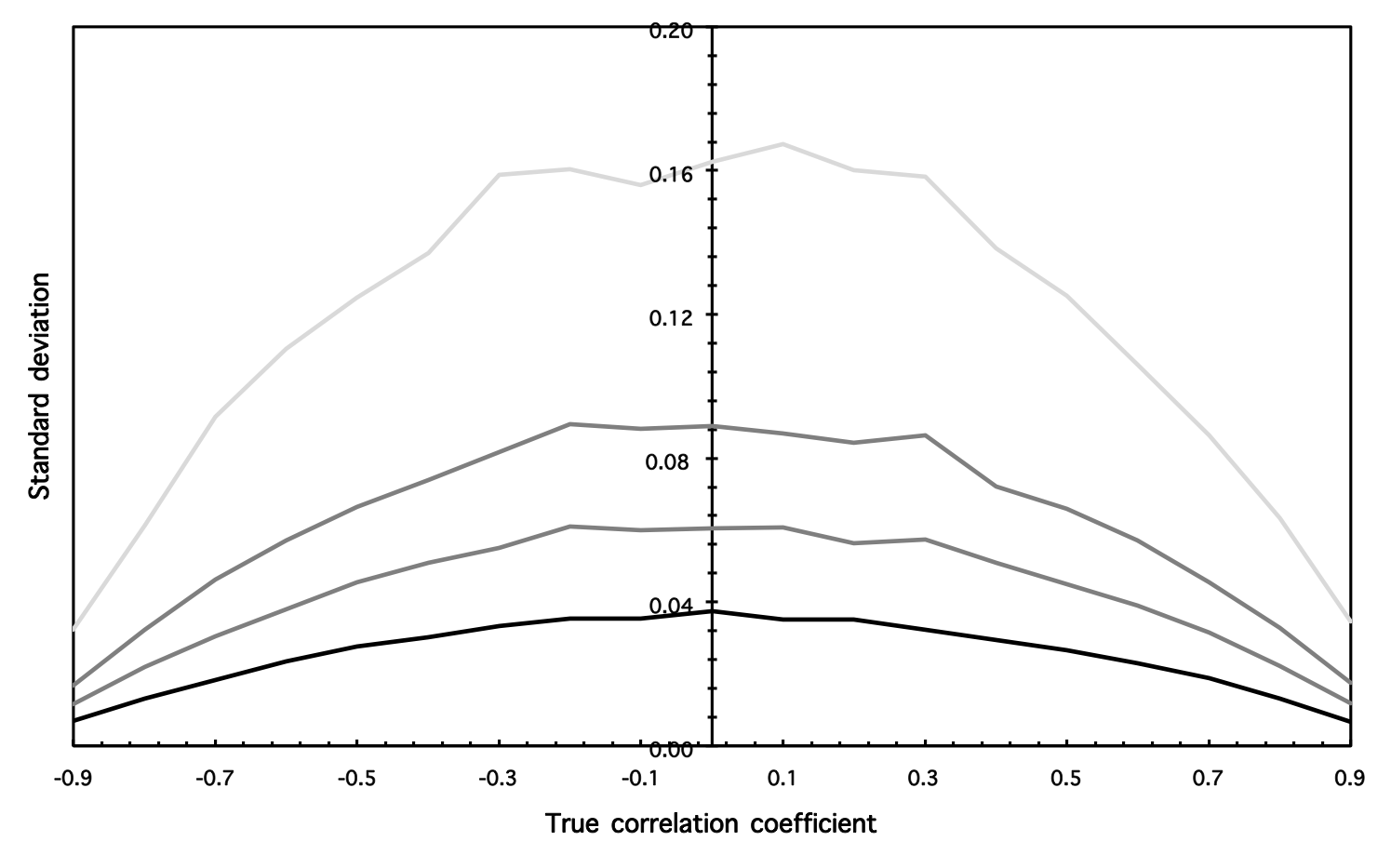}&\includegraphics[width=45mm]{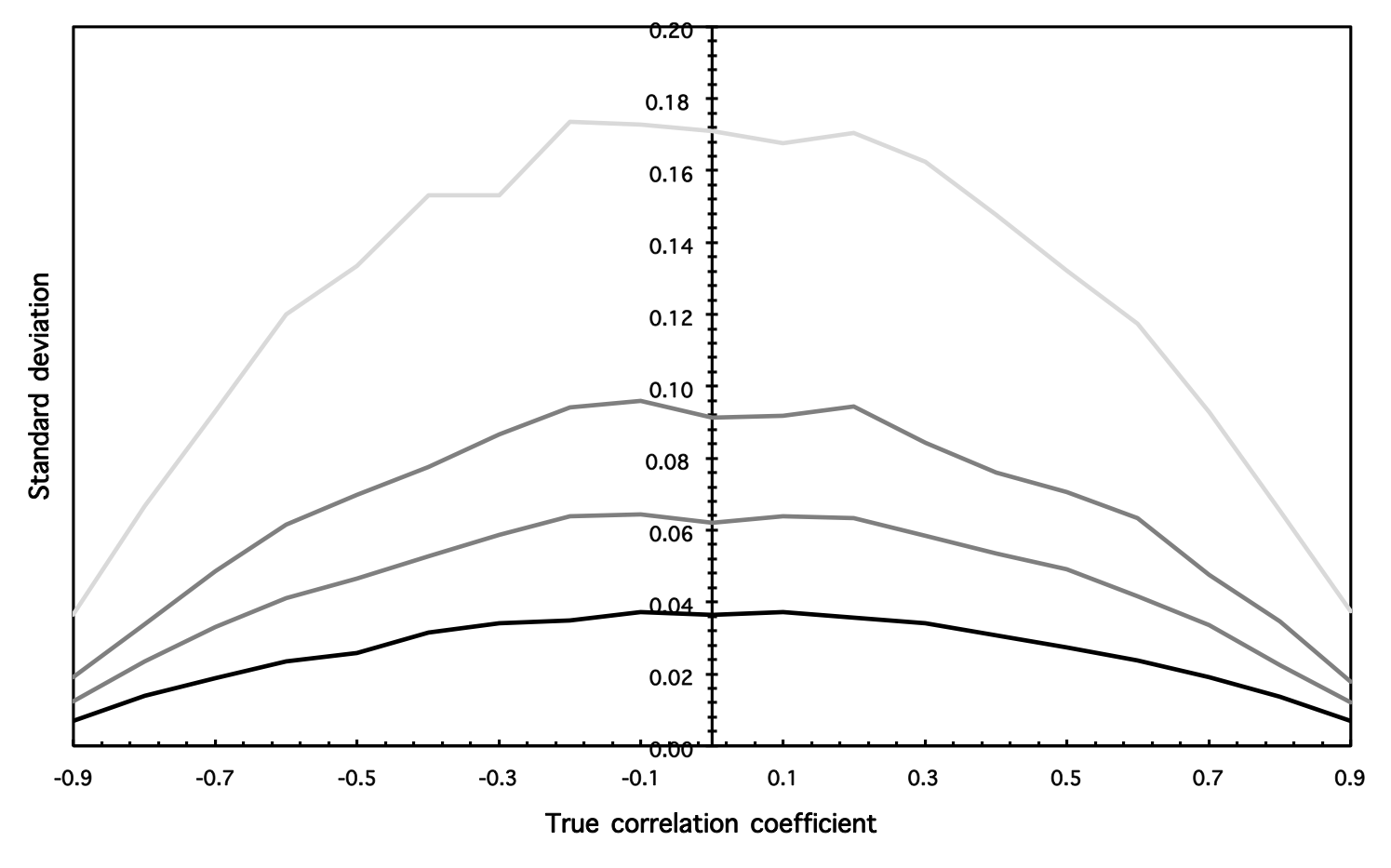}\\
\includegraphics[width=45mm]{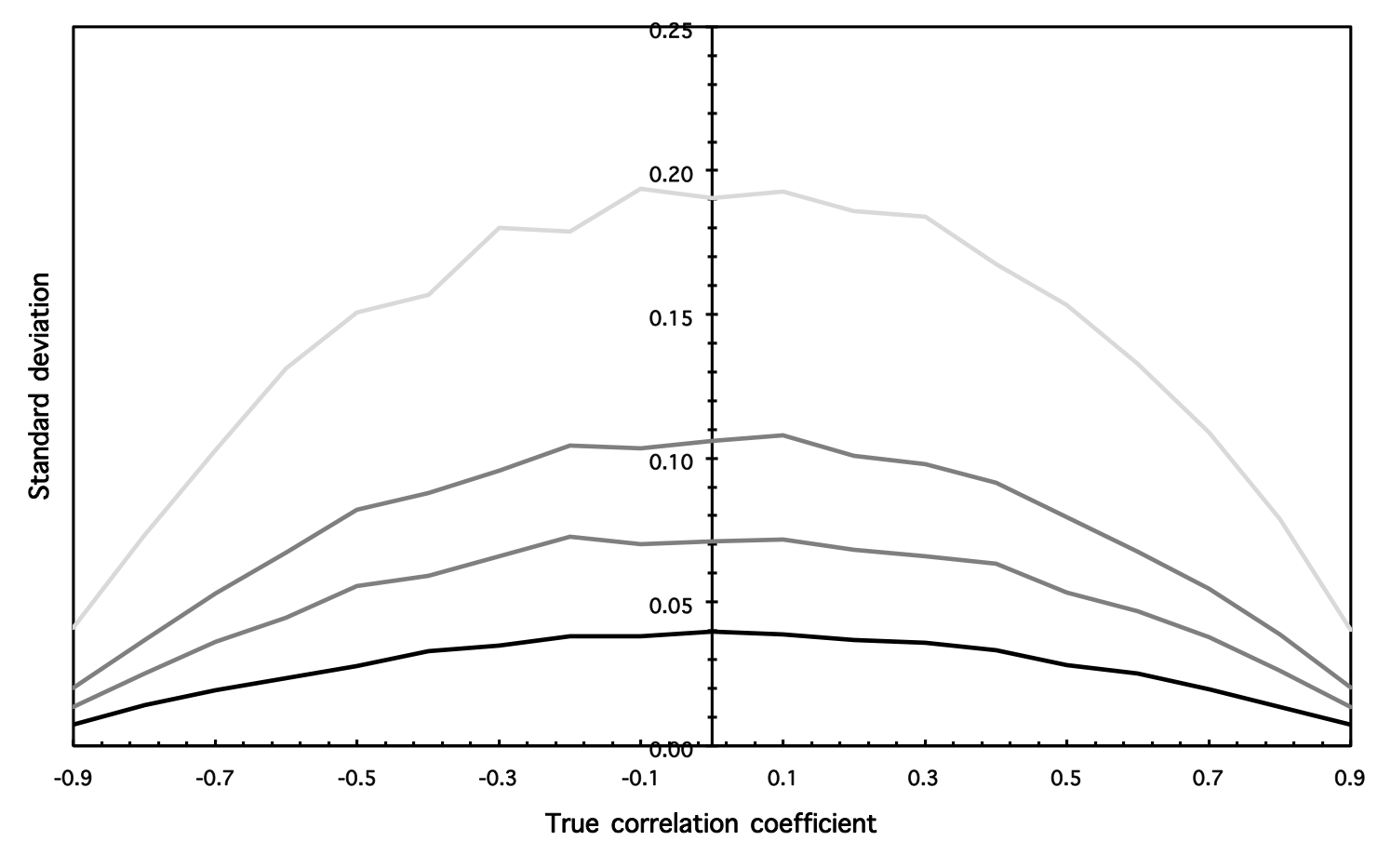}&\includegraphics[width=45mm]{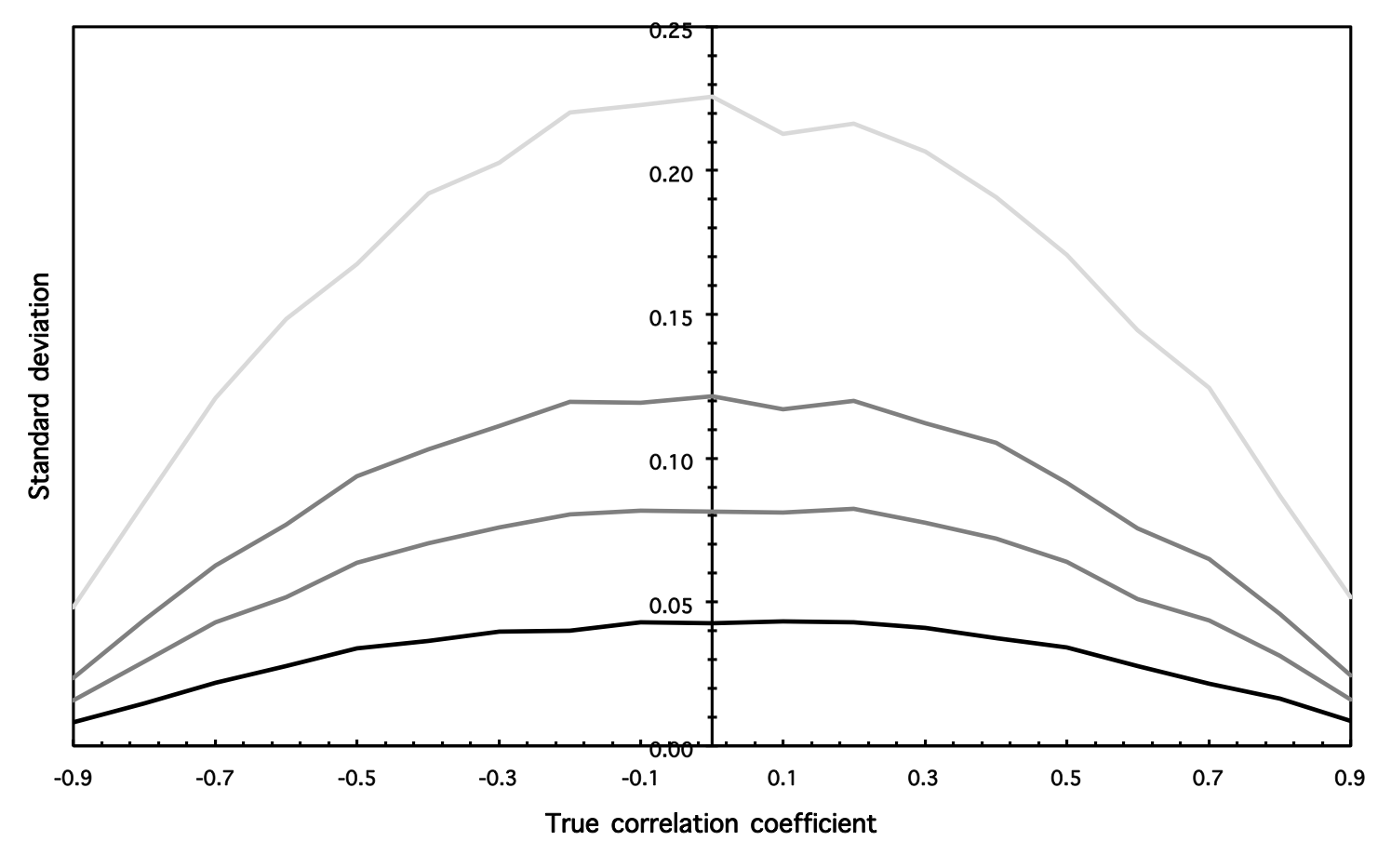}&\includegraphics[width=45mm]{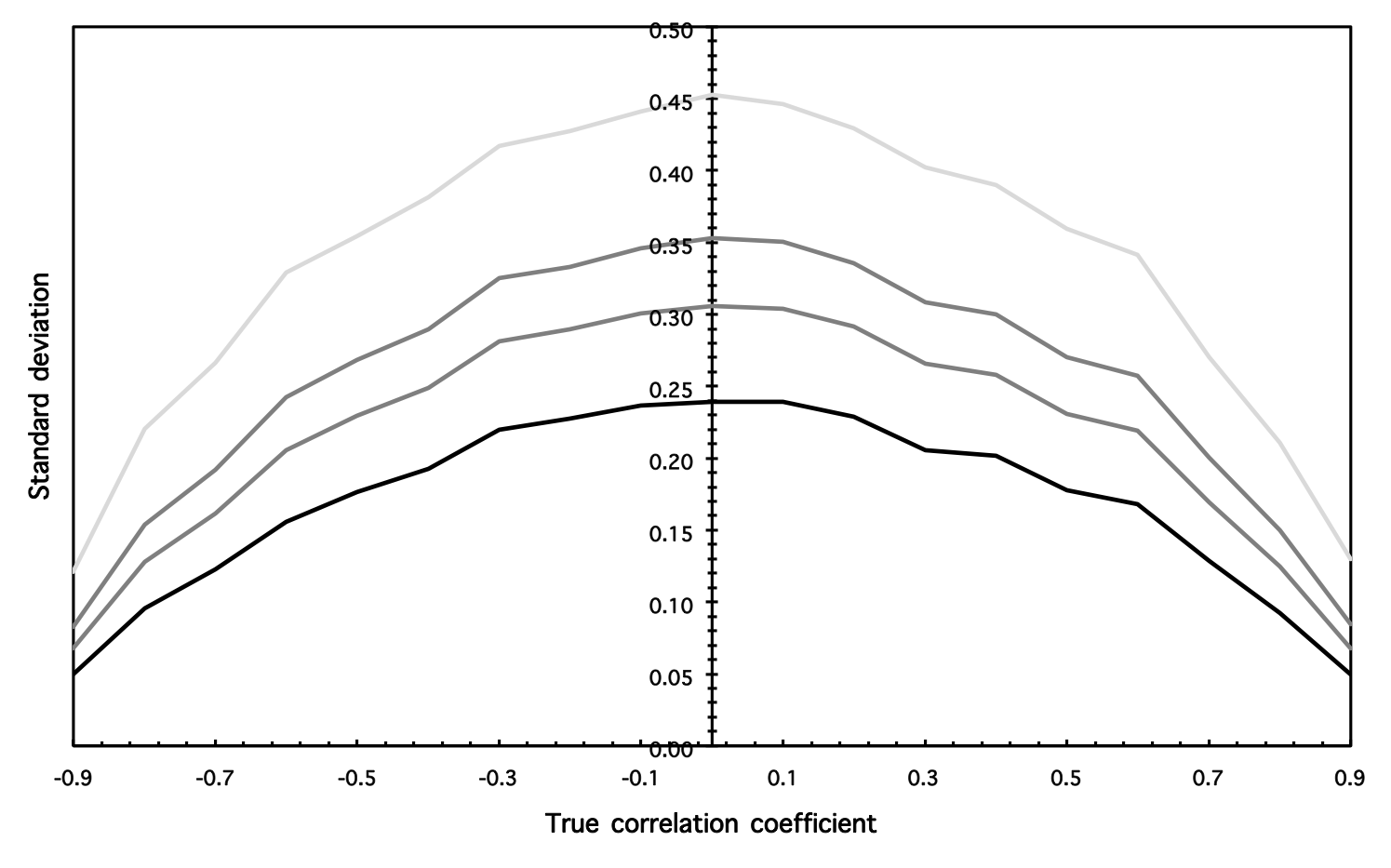}\\
\end{tabular}
\caption{\textbf{Standard deviations of DMCA correlation coefficients for different fractional integration parameters $d$ I.} \footnotesize{Results for the time series of length $T=1000$ are shown here. Separate figures represent different parameters $d$ -- $d=0.1$ (top left), $d=0.4$ (top center), $d=0.6$ (top right), $d=0.9$ (bottom left), $d=1.1$ (bottom center), $d=1.4$ (bottom right). Solid lines represent the standard deviation of 1000 simulations for given parameter setting. Different shades of grey stand for different values of the moving average window $\lambda$ going from the lowest one ($\lambda=5$, the darkest shade) to the highest one ($\lambda=101$, the lightest shade).}\label{fig2}}
\end{center}
\end{figure}

\begin{figure}[!htbp]
\begin{center}
\begin{tabular}{ccc}
\includegraphics[width=45mm]{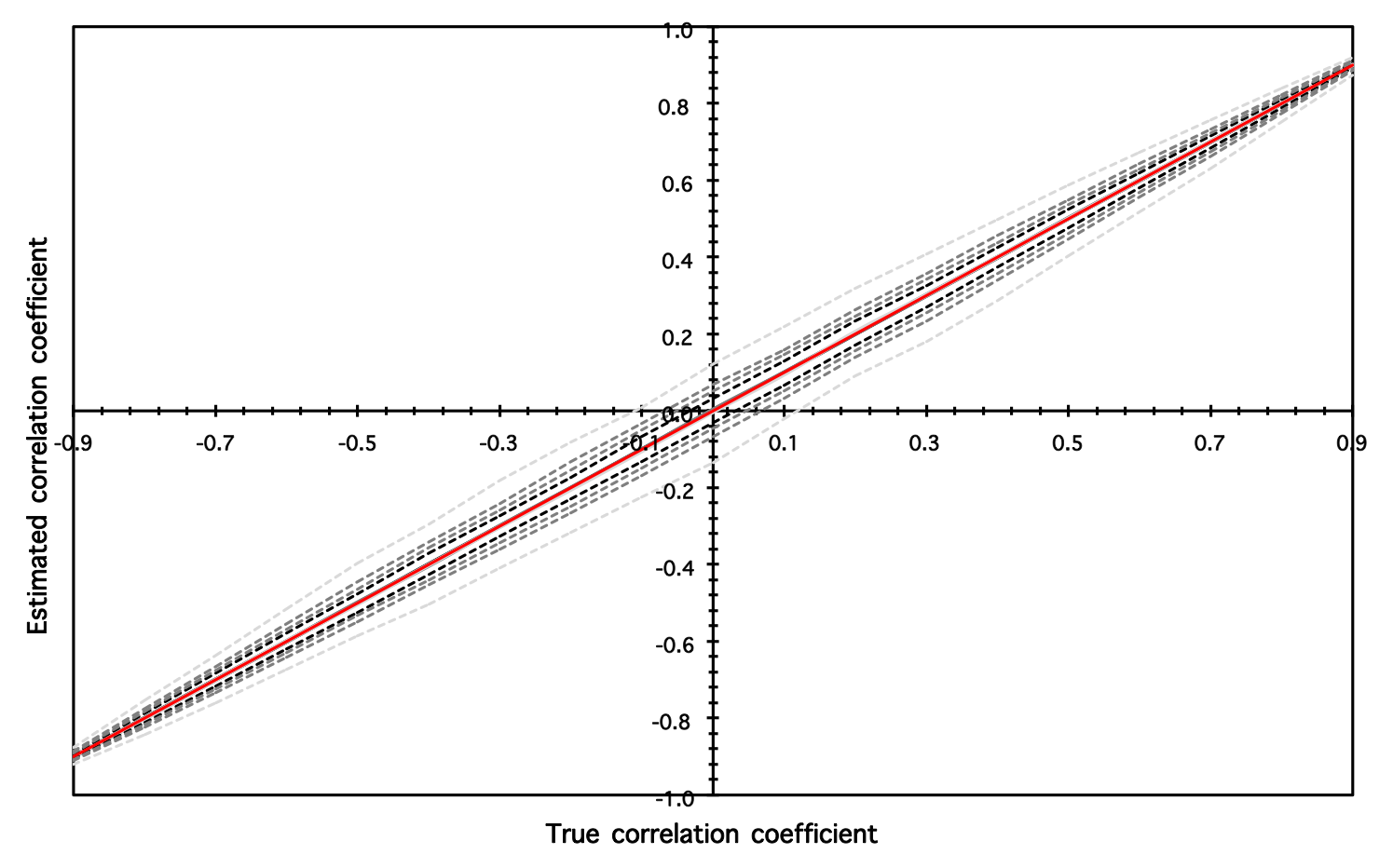}&\includegraphics[width=45mm]{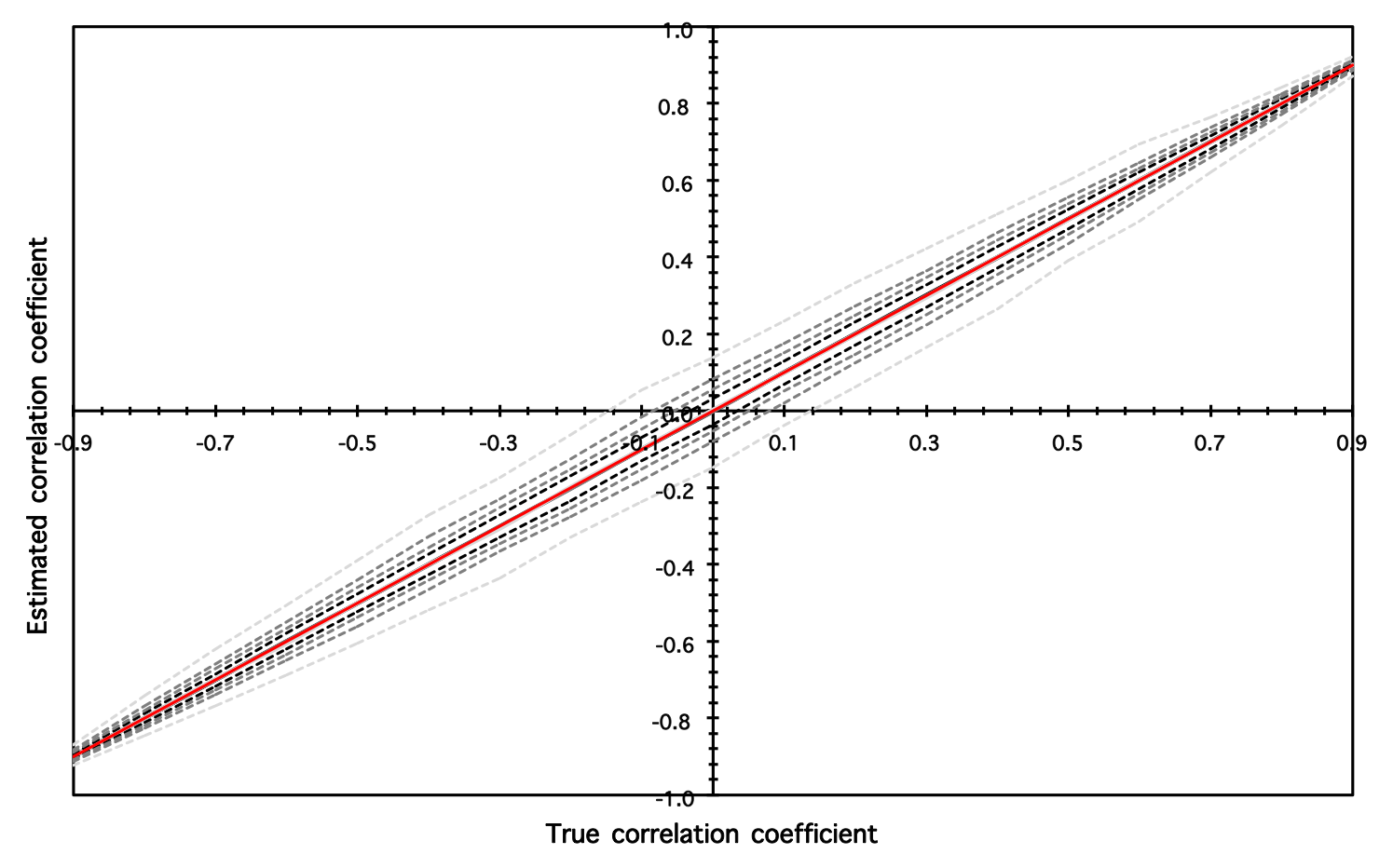}&\includegraphics[width=45mm]{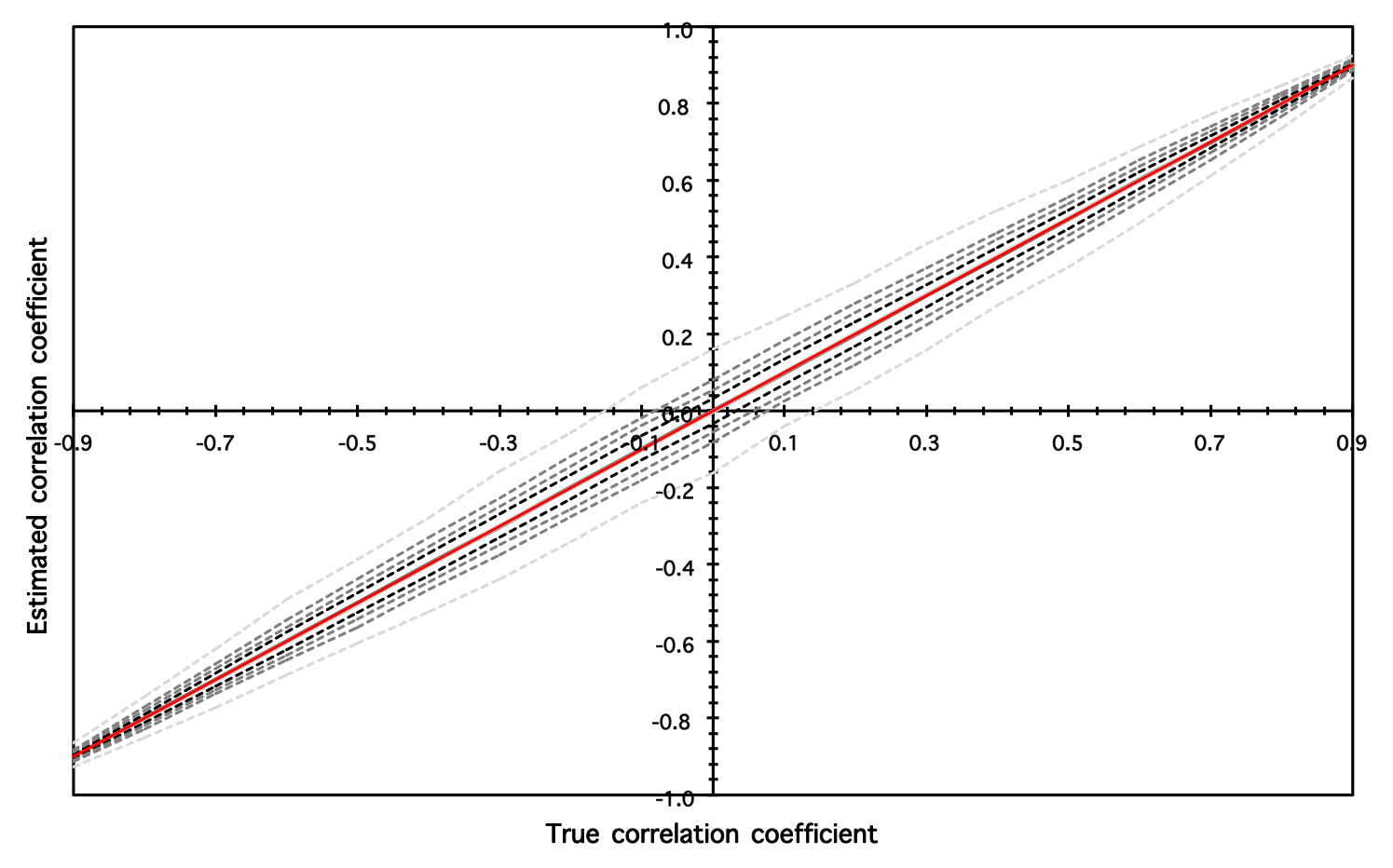}\\
\includegraphics[width=45mm]{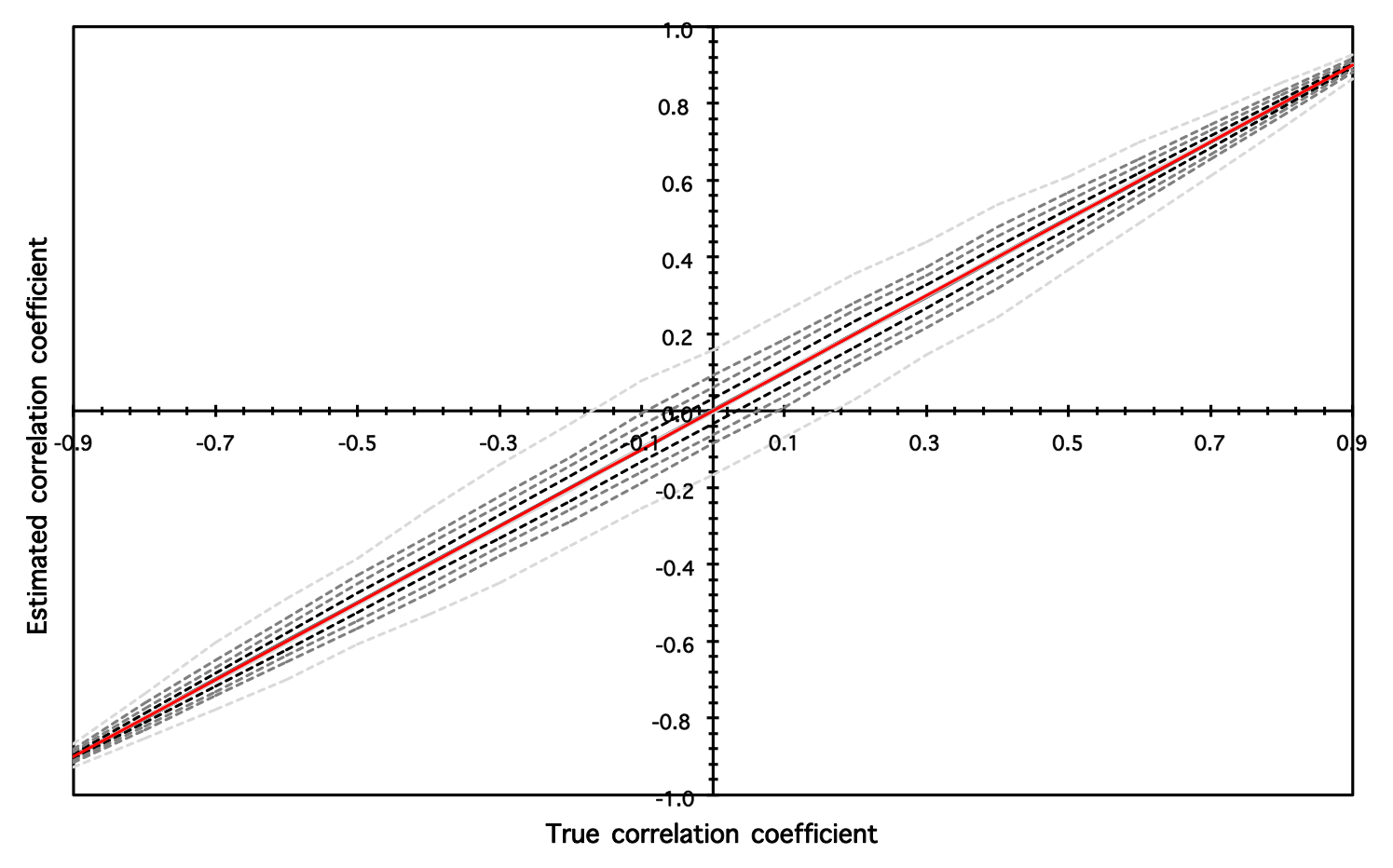}&\includegraphics[width=45mm]{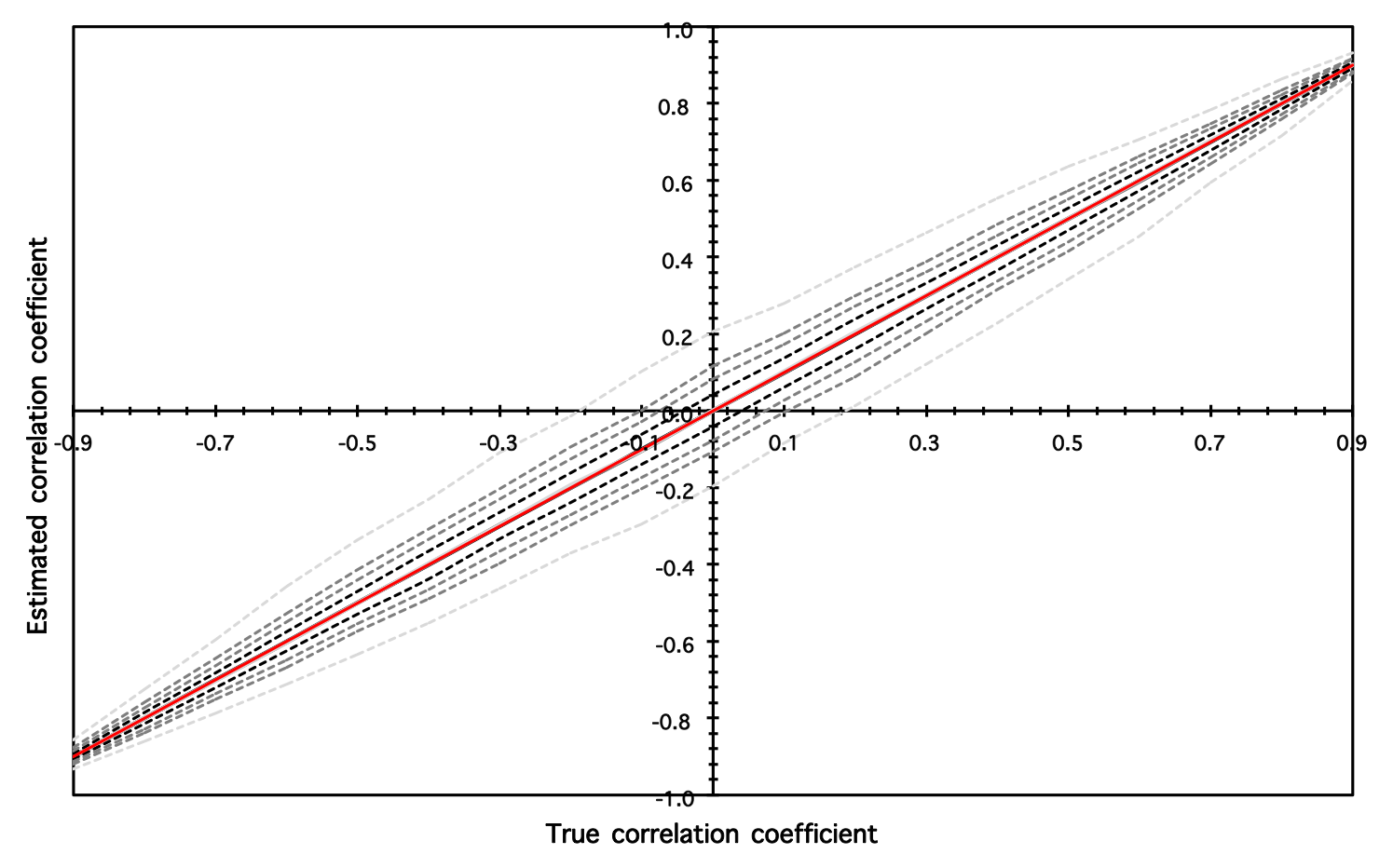}&\includegraphics[width=45mm]{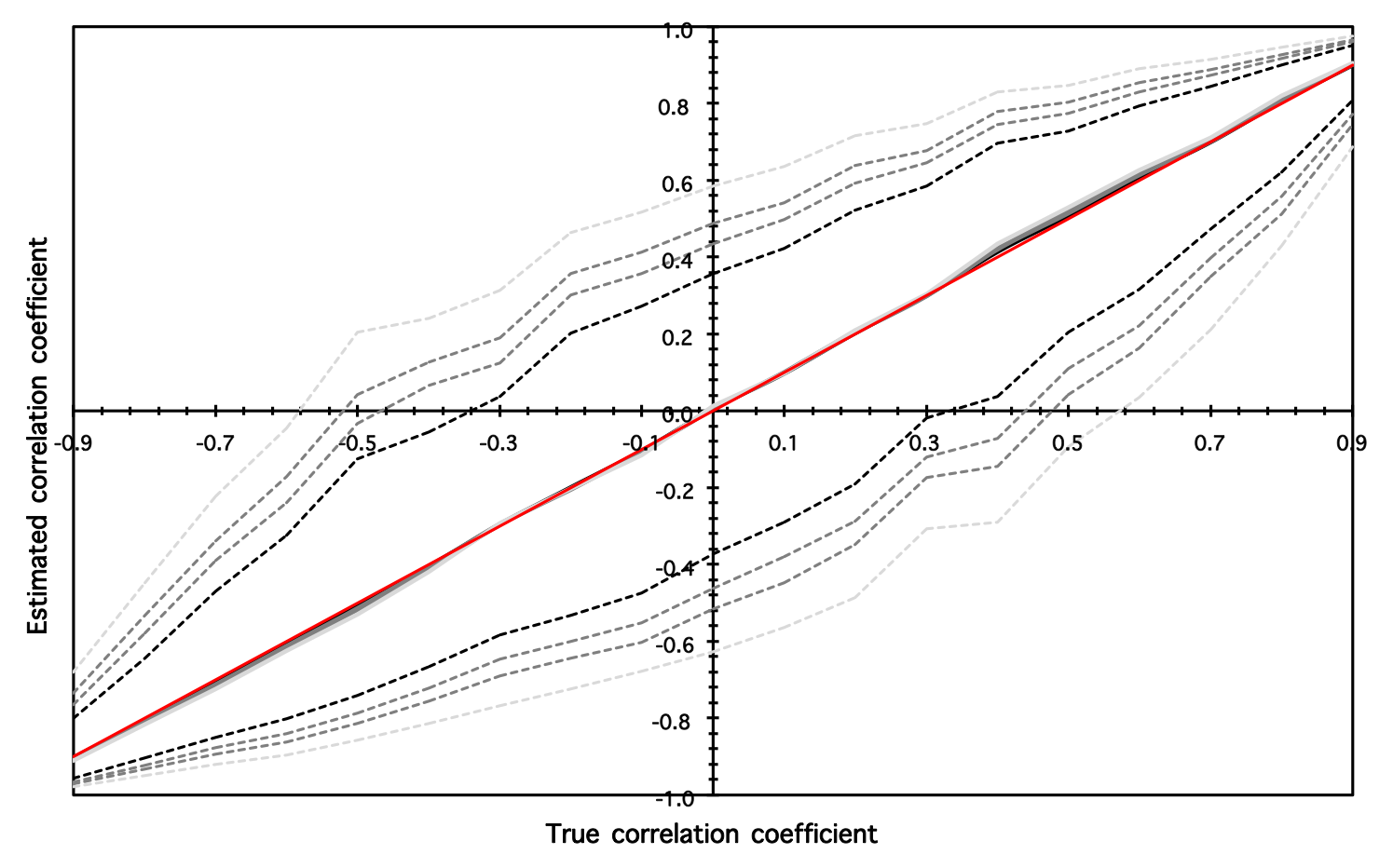}\\
\end{tabular}
\caption{\textbf{Estimated DMCA correlation coefficients for different fractional integration parameters $d$ II.} \footnotesize{Results for the time series of length $T=5000$ are shown here. Notation of Fig. \ref{fig1} is used.}\label{fig3}}
\end{center}
\end{figure}

\begin{figure}[!htbp]
\begin{center}
\begin{tabular}{ccc}
\includegraphics[width=45mm]{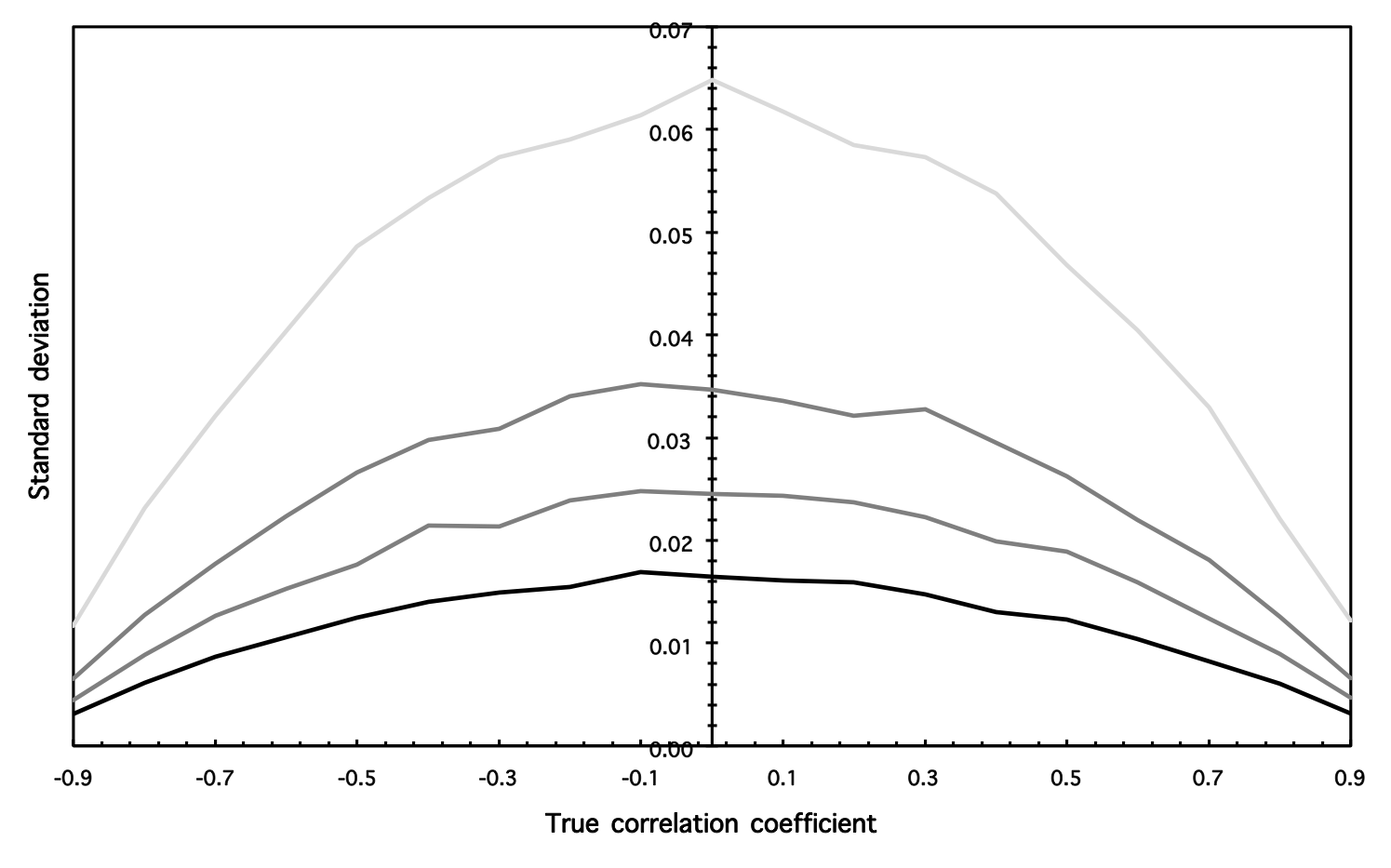}&\includegraphics[width=45mm]{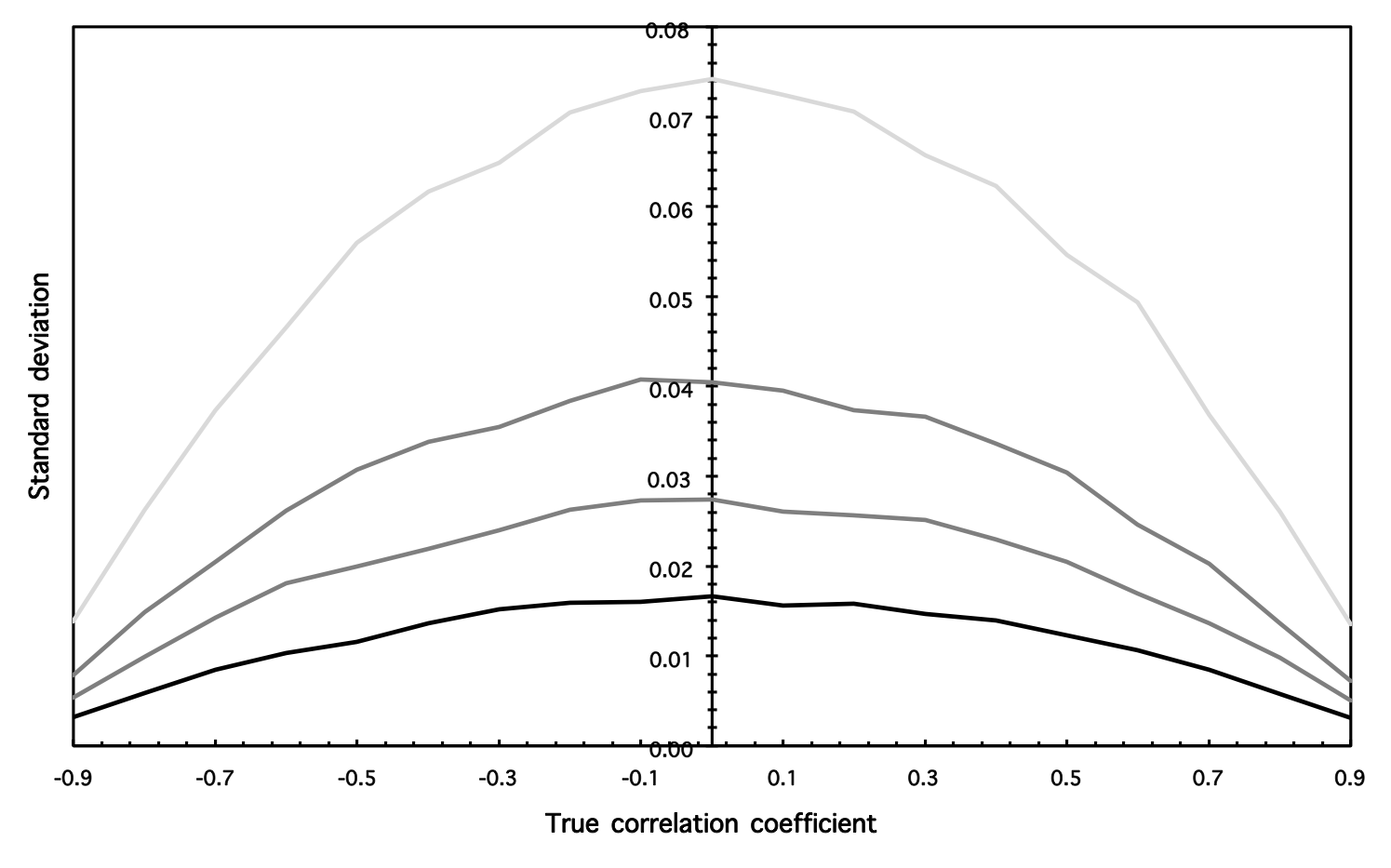}&\includegraphics[width=45mm]{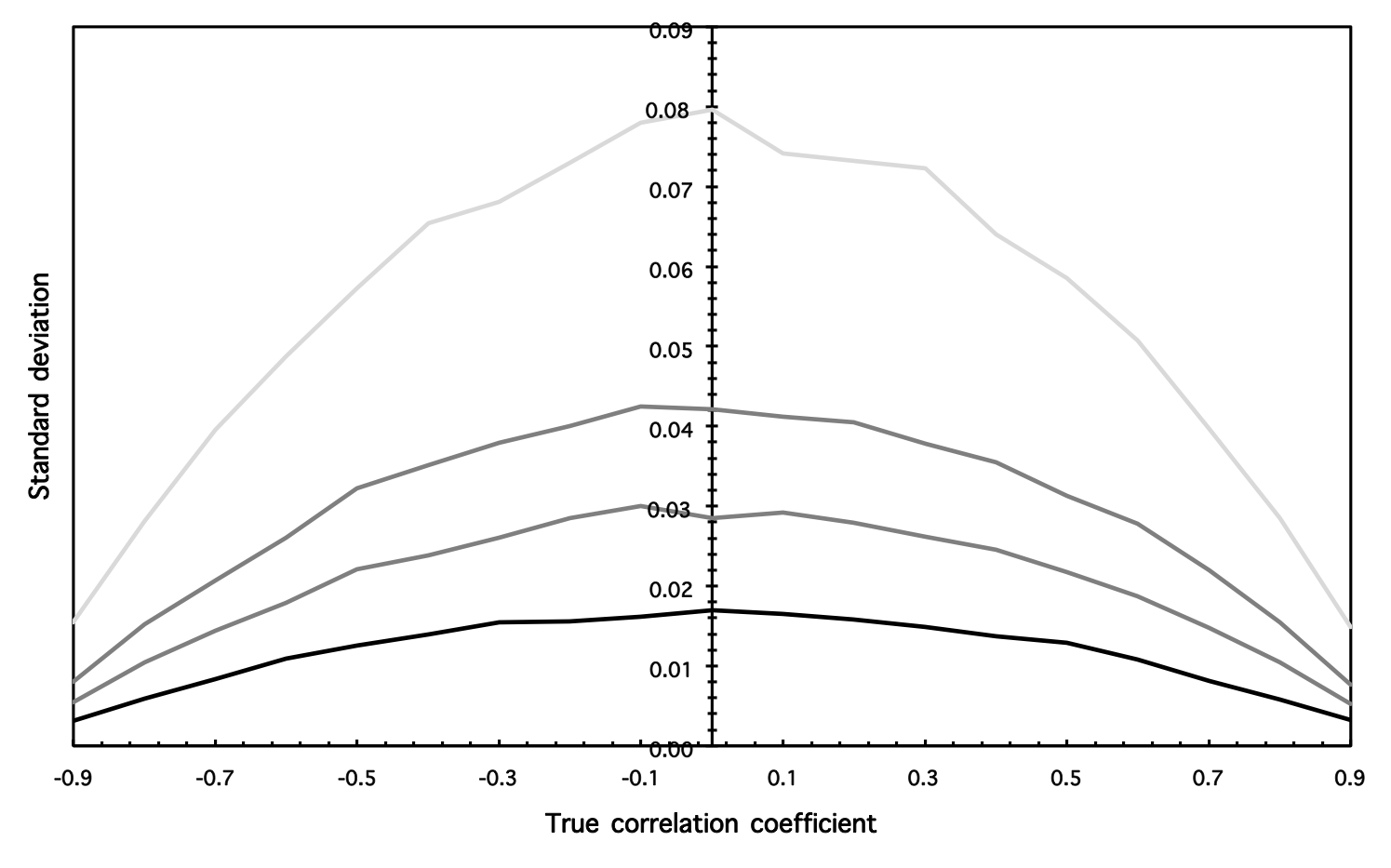}\\
\includegraphics[width=45mm]{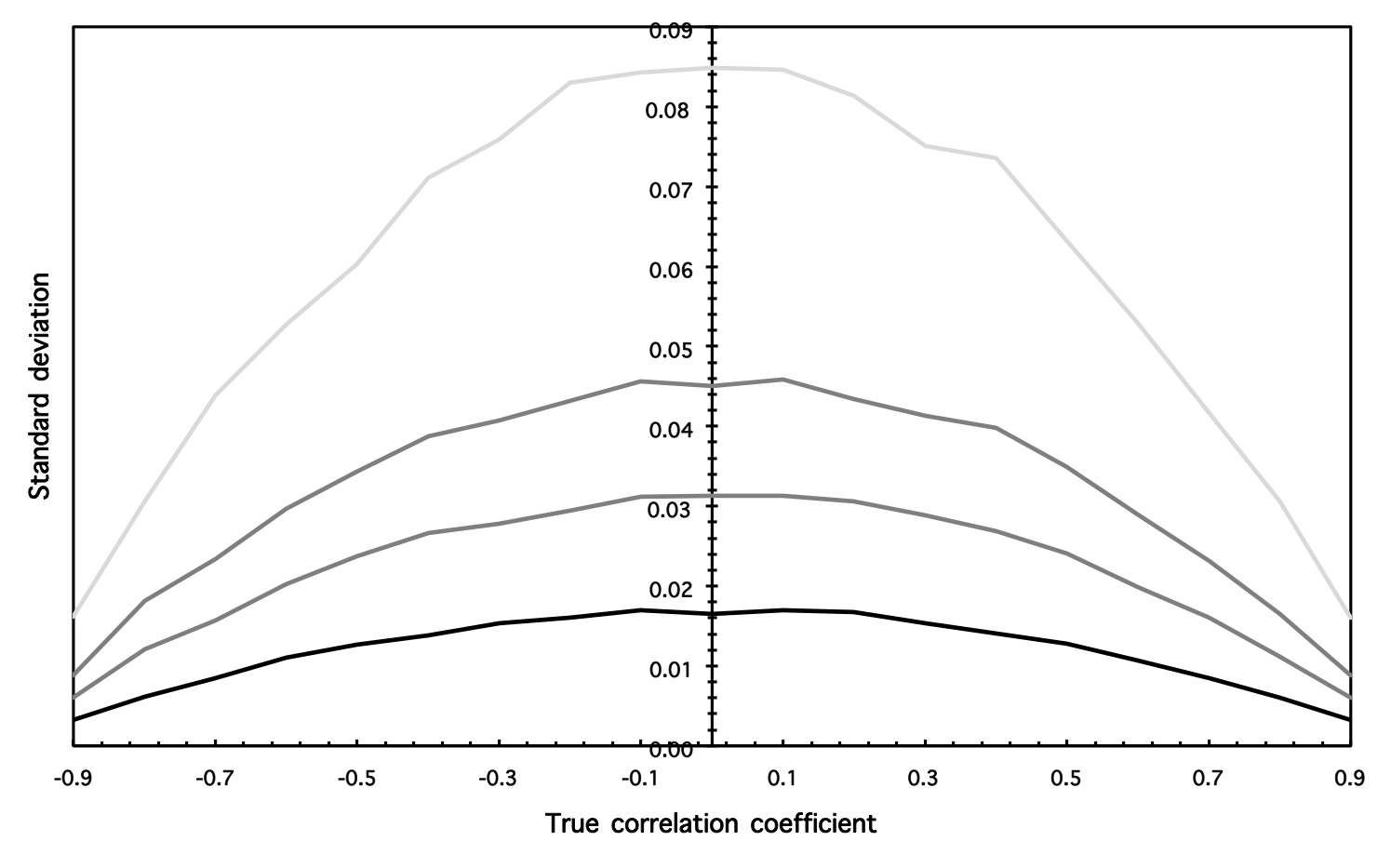}&\includegraphics[width=45mm]{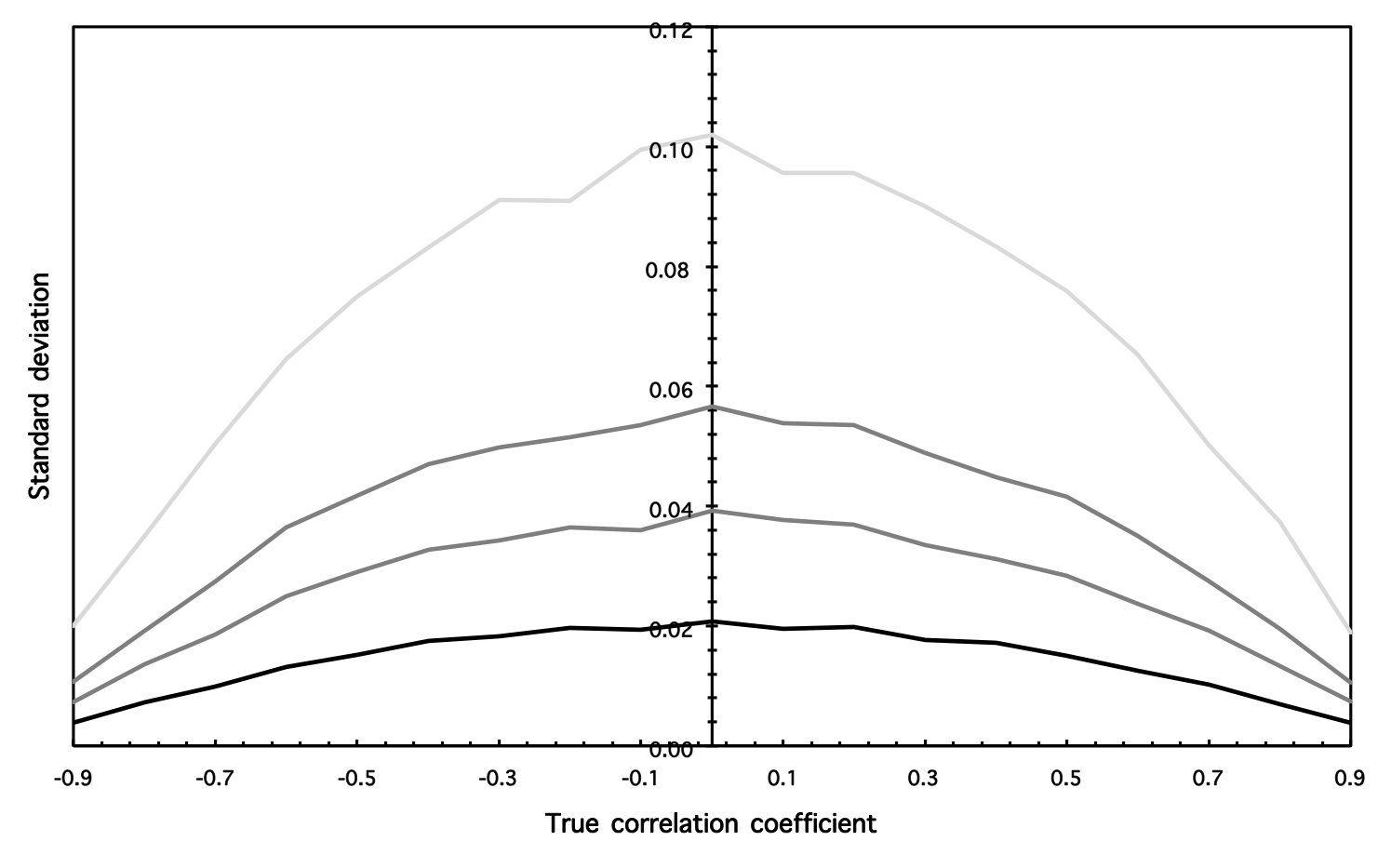}&\includegraphics[width=45mm]{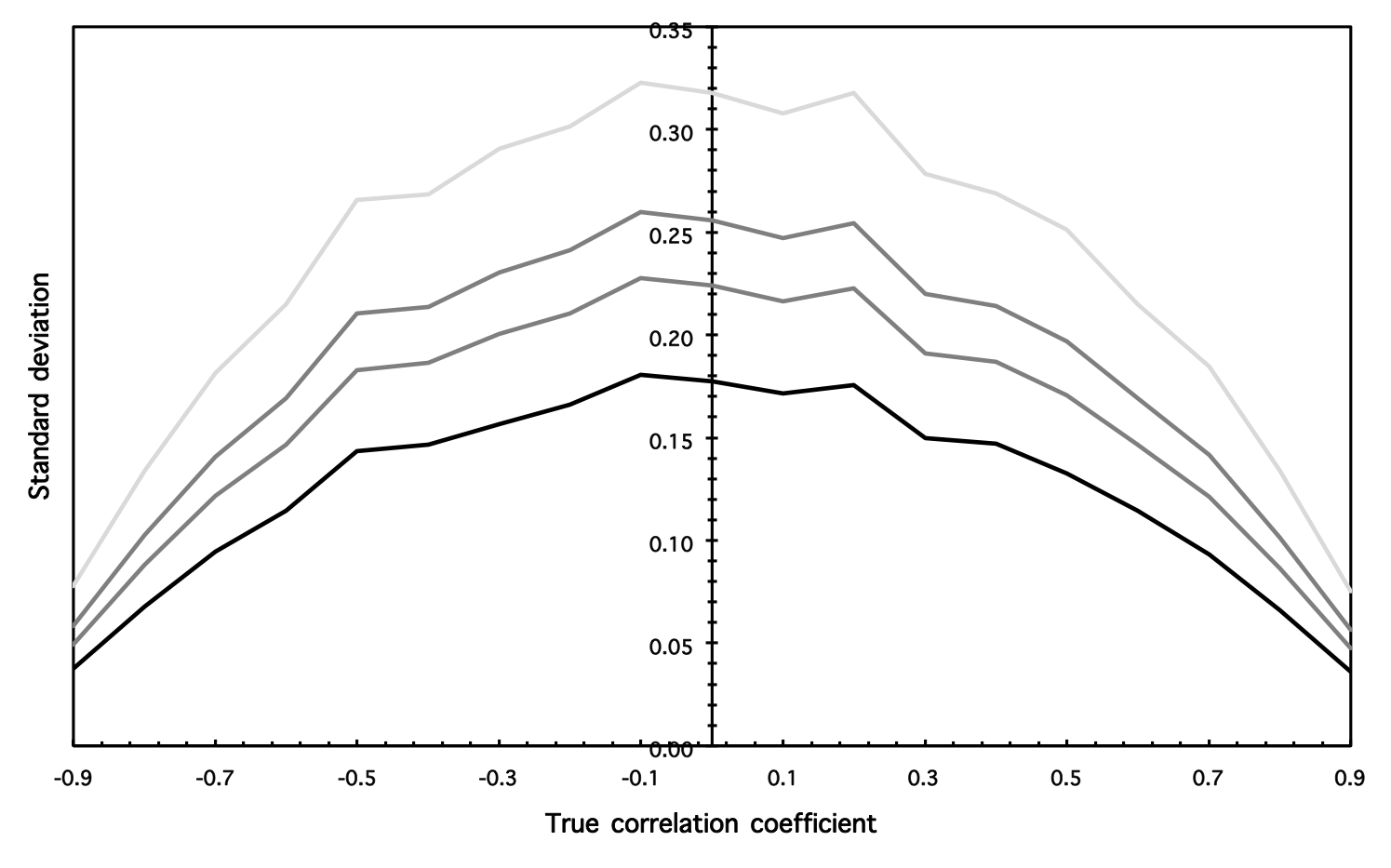}\\
\end{tabular}
\caption{\textbf{Standard deviations of DMCA correlation coefficients for different fractional integration parameters $d$ II.} \footnotesize{Results for the time series of length $T=5000$ are shown here. Notation of Fig. \ref{fig2} is used.}\label{fig4}}
\end{center}
\end{figure}

In Figs. \ref{fig1}-\ref{fig4}, all the results are summarized. In Figs. \ref{fig1}-\ref{fig2}, the time series length of $T=1000$ is discussed, and in Figs. \ref{fig3}-\ref{fig4}, the results for $T=5000$ are illustrated. In the figures, we present the 2.5\%, 50\% and 97.5\% quantiles, i.e. we show the 95\% confidence intervals and the median value, based on 1000 simulations for a given parameter setting.

The main findings can be summarized as follows. Firstly, the DMCA coefficient is an unbiased estimator of the true correlation coefficient of the series regardless the (non-)stationarity setting, the correlation level, the time series length and the moving average window size $\lambda$. Secondly, the confidence intervals get wider with an increasing $\lambda$. Nonetheless, the confidence intervals remain quite narrow for all inspected $\lambda$s (when compared to the DCCA coefficient or Pearson's correlation examined by Kristoufek \cite{Kristoufek2013b}). Thirdly, the performance of the coefficient is symmetric around the zero correlation, i.e. there are no evident difference between the performance for the positive and for the negative correlations. Fourthly, the DMCA coefficient gets more precise with an increasing absolute value of the true correlation -- for the true correlation of both $\pm 0.9$, the confidence intervals are extremely narrow. Fifthly, the standard deviation of the coefficient is approximately symmetric around the zero correlation and it increases with the parameter $d$. And sixthly, the performance of the DMCA coefficient gets better with an increasing time series length $T$.

\section{Conclusions}

In the paper, we introduce a new measure of correlation between possibly non-stationary series. As the measure is based on the detrending moving-average cross-correlation analysis (DMCA), we label it as the DMCA coefficient $\rho_{DMCA}(\lambda)$ with a moving average window size $\lambda$. We analytically show that the coefficient ranges between -1 and 1 as a standard correlation does. In the simulation study, we show that the values of $\rho_{DMCA}(\lambda)$ very well correspond to the true correlation between the analyzed series regardless the (non-)stationarity level.

As the $\rho_{DMCA}(\lambda)$ coefficient can be seen as both an alternative and a complement to the $\rho_{DCCA}(s)$ coefficient \cite{Zebende2011}, precision of these coefficients should be compared. As both this study and our previous study of the statistical properties of the DCCA coefficient \cite{Kristoufek2013b} are constructed in a similar manner, the comparison is easy. The most important findings are the following. Firstly, both coefficients provide an unbiased estimator even for highly non-stationary processes. Secondly, apart from the very strong non-stationarity case ($d=1.4$), both coefficients bring very precise estimates of the true correlation. And thirdly, turning to the differences, we find that the DMCA coefficient has narrower confidence intervals and lower standard deviations for given settings. However, it needs to be noted that the crucial parameters of the coefficients are different. For the DMCA method, we have the moving average window length $\lambda$, and for the DCCA method, we utilize the scale $s$ which is used for box-splitting and averaging of the fluctuations around the time trend. Therefore, these two coefficients are not easily comparable. However, an indisputable advantage of the newly proposed DMCA coefficient remains -- the DMCA procedure is computationally much less demanding than the DCCA method mainly due to the box-splitting and the time-trend fitting in the DCCA procedure. This might become an issue for long financial series and in any other branch of research. Overall, we suggest to use both methods in empirical analyses dealing with potentially non-stationary series and each of the methods can better control for different types of trends. 

\section*{Acknowledgements}
The support from the Grant Agency of Charles University (GAUK) under project $1110213$, Grant Agency of the Czech Republic (GACR) under projects P402/11/0948 and 402/09/0965, and project SVV 267 504 are gratefully acknowledged.

\bibliography{DMCA}
\bibliographystyle{chicago}

\newpage

\end{document}